\documentstyle[12pt]{article}
         \makeatletter
         
         \def\thefigure{\@arabic\c@figure}\def\fps@figure{tbp}
         \def\ftype@figure{1}\def\ext@figure{lof}
         \def\fnum@figure{\protect\footnotesize Fig.\ \thefigure}
         \def\thetable{\@arabic\c@table}
         \def\fps@table{tbp}\def\ftype@table{2}\def\ext@table{lot}
         \def\fnum@table{\protect\footnotesize Table \thetable}
         \def\@listI{\leftmargin\leftmargini\parsep=0pt\itemsep=0pt}
         \def\thebibliography#1{\section{References}\vspace*{-10pt}\list
          {[\arabic{enumi}]}{\settowidth\labelwidth{[#1]}\leftmargin\labelwidth
          \advance\leftmargin\labelsep
          \usecounter{enumi}}
          \def\newblock{\hskip .11em plus .33em minus .07em}
          \sloppy\clubpenalty4000\widowpenalty4000
          \sfcode`\.=1000\relax}
         
         \def\@nomath#1{\ifmmode \fi}
         \def\mmycite{\@ifnextchar [{\@tempswatrue\@mmycitex}
             {\@tempswafalse\@mmycitex[]}}
         \def\@mmycitex[#1]#2{\if@filesw\immediate%
         \write\@auxout{\string\citation{#2}}\fi
           \def\@citea{}\@mmycite{\@for\@citeb:=#2\do
             {\@citea\def\@citea{,}\@ifundefined
                {b@\@citeb}{{\bf ?}\@warning
                {Citation `\@citeb' on page \thepage \space undefined}}%
         \hbox{\csname b@\@citeb\endcsname}}}{#1}}
         \def\@mmycite#1#2{{{\scriptsize#1}\if@tempswa , #2\fi}}
         \def\mycite#1{$^{\protect\mmycite{#1}}$}
         \def\@cite#1#2{{#1\if@tempswa , #2\fi}}
         \def\thesection {\arabic{section}}
         \def\section#1{\addtocounter{section}{1}\setcounter{subsection}{0}
              \bigskip\medskip{\noindent\bf\thesection.\ #1}
              \medskip}
         \def\thesubsection {\arabic{section}.\arabic{subsection}}
         \def\subsection#1{\addtocounter{subsection}{1}
              \medskip{\noindent\thesubsection.\ #1}
              \medskip}
         \makeatother
\begin{document}
%
%
\def\deg{$^{\circ}$}
\def\cii{$^{12}$C}
\def\clvii{$^{37}$Cl}
\def\clv{$^{35}$Cl}
\def\arvi{$^{36}$Ar}
\def\mgiv{$^{24}$Mg}
\def\neo{$^{20}$Ne}
\def\sige{$\Sigma E_{kin}$}
\def\sigq{$\Sigma Q_0$}
\def\sz{$\Sigma$Z}
\vspace*{0.3in}

\begin{center}
  {\bf Signatures of Statistical Decay}\\
  \bigskip
  \bigskip
D. Horn$^a$, G.C. Ball$^a$, D. R. Bowman$^a$, A. Galindo-Uribarri$^a$,\\
E. Hagberg$^a$,
R. Laforest$^b$
\footnote
{Present address: Laboratoire de Physique Corpusculaire, ISMRA et
Universit\'{e} de Caen, Blvd. du Mar\'{e}chal Juin, F-14050 Caen,
France.},
J. Pouliot$^b$
\footnote
{Present address: H\^otel-Dieu de Qu\'ebec, D\'epartement de
Radio-Oncologie, Qu\'ebec, Canada.},
and R. B. Walker$^a$\\

$^a${\em AECL, Chalk River Laboratories, Chalk River, Ontario,
K0J~1P0, Canada}\\
$^b${\em Laboratoire de Physique Nucl\'{e}aire, Universit\'{e} Laval,
Ste-Foy, Qu\'{e}bec, G1K~7P4, Canada}\\

\bigskip
\end{center}
\smallskip
{\footnotesize
\centerline{ABSTRACT}
\begin{quotation}
\vspace{-0.10in}
The partition of decay energy between the kinetic energy of reaction
products and their Q-value of formation is obtained in a statistical
derivation appropriate to highly excited nuclei, and is shown to be in
a constant ratio.  We measure the kinetic energy fraction,
$R = \Sigma E_{kin}/(\Sigma E_{kin} + \Sigma Q_0)$, over a wide range of
excitation energy for well-defined systems formed in the \clv +\cii\
reaction at 35{\em A} MeV. Relationships between excitation energy,
charged-particle multiplicity, and intermediate-mass-fragment
multiplicity,observed in this work and in recent experiments by a number of
other
groups, follow from the derivation of the average kinetic energies and
Q-values.
\end{quotation}}

\section{Introduction}

A number of scaling phenomena and correlations between observables have
recently been discovered in the deexcitation of highly excited nuclei.
These include:
\begin{itemize}
\item the correlation of $N_{IMF}$, the number of intermediate-mass
fragments, with $N_c$, the total number of charged products,\mycite{bowm}
\item the correlation of $N_{IMF}$ with $Z_{bound}$, the total amount of
charge contained in fragments heavier than hydrogen,\mycite{hube}
\item the approximate proportionality of $N_{IMF}$ to $E_t$, the measured
transverse energy,\mycite{phai}
\item the linear relationship between $N_c$ and $T$, the nuclear
temperature,\mycite{pori} and
\item the scaling of IMF multiplicity yield ratios with
$\frac{1}{\sqrt{E_t}}$. \mycite{mor1,mor2}
\end{itemize}

\begin{figure}[tbh]
\vspace*{3.0 in}
\unitlength 1cm
\includegraphics{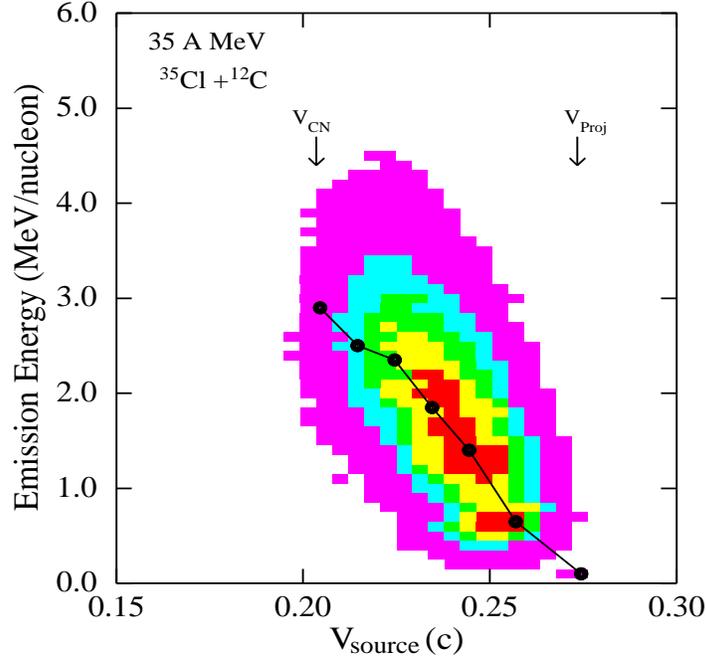}
\vspace{1.2 cm}
\caption{\protect\footnotesize Emission energy ($E_{kin}+Q_0$) per
nucleon in the moving frame plotted as a function of the velocity of the
center of mass of the detected fragments for 35$A$ MeV \clv+\cii. Points
represent a massive-transfer simulation, filtered by the experimental
acceptance; arrows indicate the velocities of the projectile and of the
projectile-target center of mass.}
\end{figure}

Each of these observations has been taken as indicative of statistical
decay. We here introduce a schematic derivation for the partitioning of
the decay energy between the kinetic and the mass-excess degrees of
freedom. Because of its relatively small width at moderate multiplicities,
the ratio of the kinetic energy of the products to the total available
decay energy is proposed as an event-by-event signature for statistical
decay. Further consequences of the derivation include the dependence of
the mean multiplicities, $<N_c>$ and $<N_{IMF}>$, on excitation energy
and the prediction of multiplicity yield ratios. The derived quantities
are confronted with our experimental results for the decay of light
nuclear systems with measured velocities, sizes, and excitation energies.

\section{Derivation of Statistical Observables}

The probability of emitting a particle with kinetic energy, $E_{kin}$,
from a state of excitation energy, $E$, is
\begin{equation}
P(E,E_{kin},Q_0) \propto\ E_{kin} \ \sigma _{inv} (E_{kin}) \ \frac
{\rho _f (E-E_{kin}-Q_0)}{\rho _i (E)},
\end{equation}
where $\rho _f$ and $\rho _i$ are the final- and initial-state level
densities and $\sigma _{inv} (E_{kin})$ is the cross section for the
inverse reaction.
The mean kinetic energy may be derived by integrating $E_{kin}$ with the
probability distribution of equation (1):
\begin{equation}
<E_{kin}> = \frac {\int _0^{E-Q_0} E_{kin} \ P(E_{kin}) dE_{kin}}
{\int _0^{E-Q_0} P(E_{kin}) dE_{kin}}.
\end{equation}
In the limit of  high excitation energy and negligible emission barriers,
this gives the traditional\mycite{morr} results for the mean kinetic
energy,
\begin{equation}
<E_{kin}> = 2\sqrt{E/a} = 2T,
\end{equation}
and its variance,
\begin{equation}
\sigma_{E_{kin}}^2 = 2E/a = 2T^2,
\end{equation}
where $a$ is the nuclear level-density parameter. These results are
appropriate for neutron emission and for charged-particle emission from
light systems, where Coulomb barriers are very low. The many exit
channels accessible to states of high excitation energy permit one to
approximate the density of available ground-state Q-values as a
continuous function, $f(Q_0)$.
A similar integral may then be performed for $Q_0$, also giving a
proportionality to $\sqrt{E/a}$, with the proportionality constant
depending on $f(Q_0)$. Thus,
\begin{equation}
<Q_0> = 2\sqrt{E/a} = 2T
\end{equation}
if the density of available exit channels is a linear function of $Q_0$.
For systems in the mass range studied experimentally here, this is a
reasonable approximation. Whatever the proportionality factor, the
kinetic energy fraction,
\begin{equation}
R = \frac {\Sigma E_{kin}}{(\Sigma E_{kin} + \Sigma Q_0)},
\end{equation}
should then be a constant, independent of excitation energy, with the
specific case of $<Q_0> = 2T$, giving $<R>=0.50$.

For statistical processes, the observed width depends on the number of
samplings of the parent distribution. In our case the number of samplings
is the number, $N$, of detected charged particles. For $<R>=0.50$, that
width is
\begin{equation}
\sigma_R = \frac{1}{4\sqrt{2N}}.
\end{equation}

We now examine the relative {\em rates} for barrier-dominated and
barrier-independent emission. If the probability per unit time for
emitting $n$ particles is approximately proportional to
$exp(-n<\Delta E>/T)$, where $<\Delta E>$ is the average deexcitation
energy per particle emitted, then the particle emission {\em rate} is
\begin{equation}
<n> = \frac{\int_{}^{} dn \ n \ exp(-n\Delta E/T)}
{\int_{}^{} dn  \ exp(-n\Delta E/T)}
\propto\ \frac{T}{\Delta E}.
\end{equation}
For light-ion emission from a light, highly excited nucleus, we neglect
the Coulomb barrier and obtain
\begin{equation}
<\Delta E> \ = \ <E_{kin}> + <Q_0> \ = \ 4T.
\end{equation}
However, for emission of heavier fragments, barrier effects may dominate,
so that $<\Delta E> \approx\ B$.
A fixed or generic value of $B$, previously demonstrated\mycite{mor1} to
be appropriate for IMF emission, would imply the rates for the two
processes to be in the ratio,
\begin{equation}
\frac{<n_{IMF}>}{<n>} \propto\ \frac{T/B}{T/4T} \propto\ T.
\end{equation}
If the average deexcitation per particle is $<\Delta E>=4T$, the mean
number of particles emitted would be $<N>=E/4T$, giving
\begin{equation}
N \propto\ \sqrt{E},
\end{equation}
and variance,
\begin{equation}
\sigma_N^2 = N/4.
\end{equation}
{}From the ratio of rates,
\begin{equation}
N_{IMF} \propto\ E.
\end{equation}
If the proportionality of transverse energy to excitation energy is
assumed, then equations (11) and (13) consolidate the first four points
of the introduction.

\begin{figure}[tb]
\vspace*{3.0 in}
\unitlength 1cm
\begin{minipage}[t]{2.9in}
\centerline{
\includegraphics{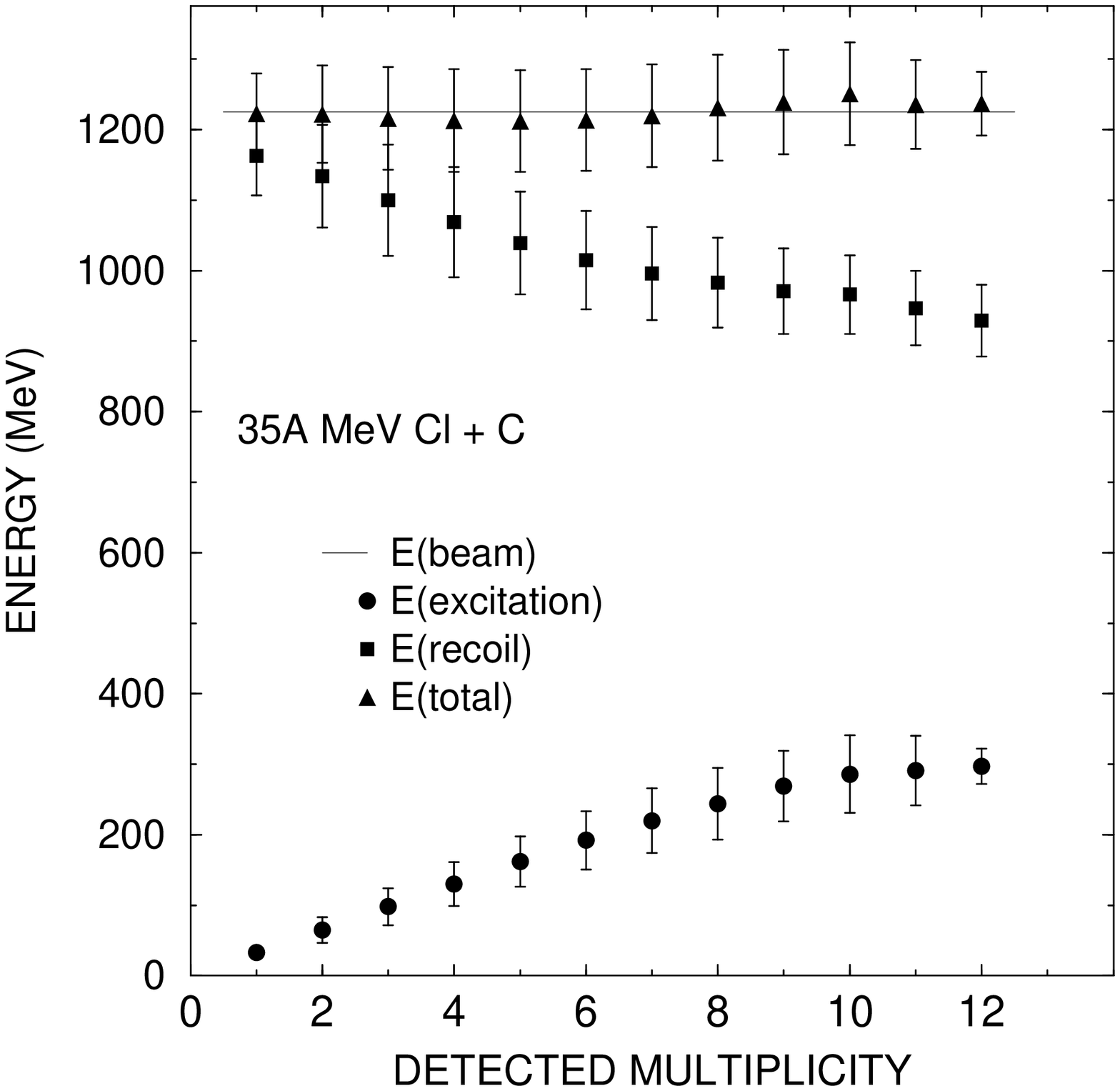}
}
\vspace{1.5 cm}
\caption{\protect\footnotesize Energy sums and distribution widths (bars)
for 35$A$ MeV \clv+\cii\ as a function of the number of detected charged
particles. The solid line indicates the beam energy.}
\end{minipage} \ \hspace{0.0in} \
\begin{minipage}[t]{2.9in}
\centerline{
\includegraphics{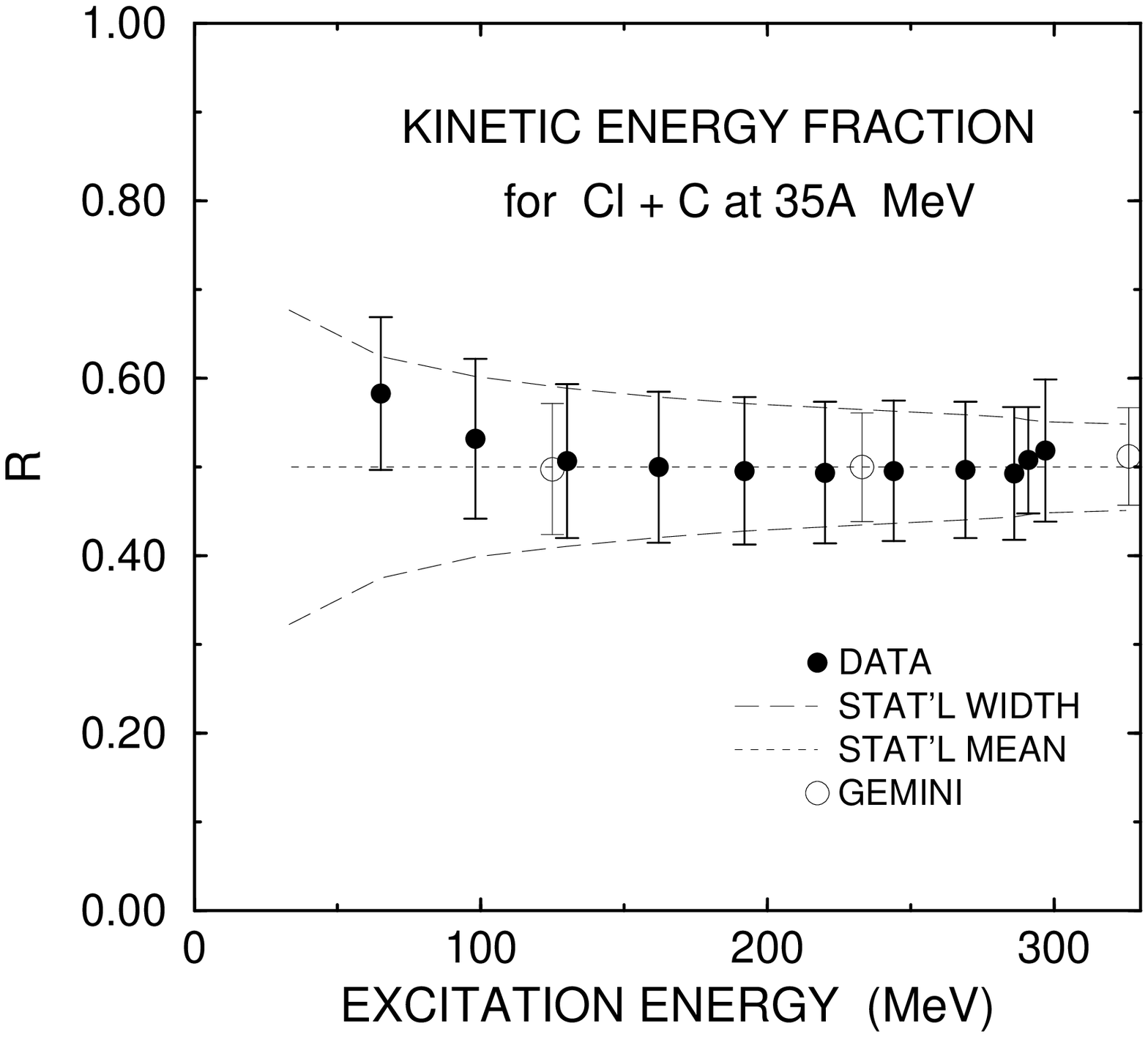}
}
\vspace{1.5 cm}
\caption{\protect\footnotesize Centroids and widths of kinetic energy
fraction, $R$, for \clv\ + \cii\ at 35$A$ MeV. Data for multiplicities
2 to 12 are plotted.}
\end{minipage}
\end{figure}

\newpage

\section{Experimental Determination of Source Properties}

A systematic comparison of the derived quantities with experiment
requires the event-by-event determination of the size, velocity and
excitation energy of the decaying system. One type of reaction  which
lends itself to the isolation and measurement of an excited source is the
massive transfer mechanism\mycite{siem}.
A beam of 35$A$ MeV \clv\ ions, provided by Chalk River's TASCC facility,
collided with a 2-mg/cm$^2$ carbon target, producing a variety of
reaction products. The kinematics of the reaction served to focus the
reaction products into the 6\deg\ to 25\deg\ angular range of our
detector array, a close-packed assembly of 40 phoswich
counters\mycite{pru1}. All ions were identified by atomic number and
isotopic distributions were measured with a set of high-resolution Si/CsI
detector telescopes.\mycite{hor1} Our selection of massive transfer
events was facilitated by the thresholds and limited angular acceptance
of the array, which reduced our sensitivity to target-like ``spectator''
matter.

To ensure that no major component of an event went undetected, we
analyzed only events in which at least 15 units of charge were
identified. The velocity of the center of mass of all detected ions was
computed, and the kinetic energy of each product was calculated within
the moving frame. Based on the average mass excess deduced from the
isotope distributions for each detected ion, $\Sigma Q_0$ was added to
$\Sigma E_{kin}$ for each event, and the resulting deexcitation energy
per detected nucleon was plotted as a function of source velocity.
Fig. 1 shows the relationship between emission energy per nucleon and
source velocity, starting at projectile-like values of source velocity
and zero emission energy in the lower right portion of the figure, and
extending to compound-nucleus velocity and large emission energy in the
upper left. The entire range of massive transfer phenomena, as previously
observed at lower energies\mycite{colo}, is evident in the figure.
The points superimposed on the figure represent centroids of the
distributions in energy and velocity  for the simulated massive transfer
of zero to twelve nucleons from the carbon target to the heavier
projectile. Decay of the excited system is simulated\mycite{hor2} by a
Monte Carlo event generator and filtered by the experimental acceptance.
Quantitative agreement with the data has been demonstrated\mycite{hor3}
for simultaneous projection of the same set of calculated results on both
the energy and velocity axes.

A massive transfer reaction has the useful feature that the mass of the
recoiling system may be deduced from its velocity, allowing a
model-dependent estimate of total excitation energy. The validity of this
estimate can then be tested by the consistency check demonstrated in
Fig. 2, where the excitation  energy, recoil energy, and deduced total
energy are plotted as a function of the number of detected charged
particles. Here, the multiplicity-dependence of the momentum transfer and
energy deposition is evident. Significantly, the energy sums yield the
initial projectile energy, indicating that the efficiency corrections
have been properly applied. The behavior of any unobserved component must
then be consistent with that of the detected fragments.

\begin{figure}[tb]
\vspace*{3.0 in}
\unitlength 1cm
\includegraphics{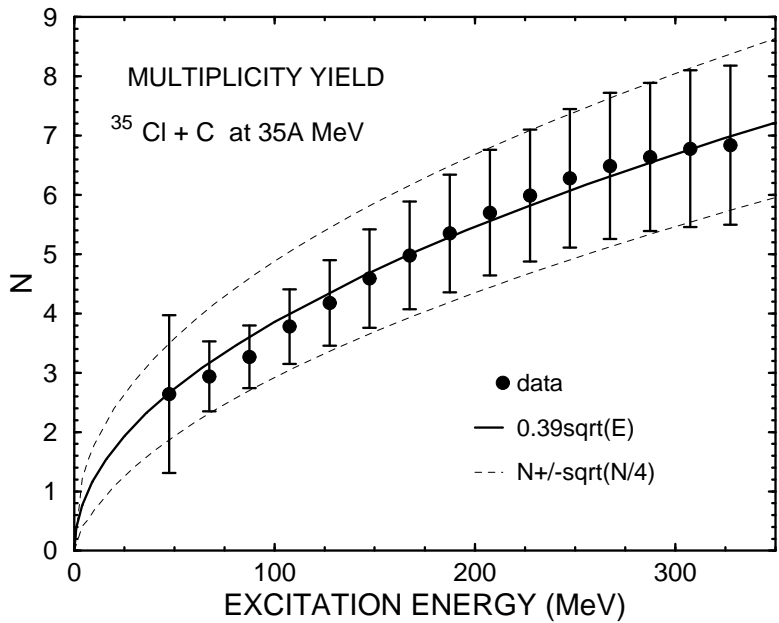}
\vspace{1.0 cm}
\caption{\protect\footnotesize Centroids and widths of multiplicity
distributions for \clv\ + \cii\ at 35$A$ MeV, plotted as a function of
reconstructed excitation energy. The proportionality constant for the
solid line, $k=0.39$, was chosen to fit the data.}
\end{figure}

\begin{figure}[tb]
\vspace*{3.0 in}
\unitlength 1cm
\includegraphics{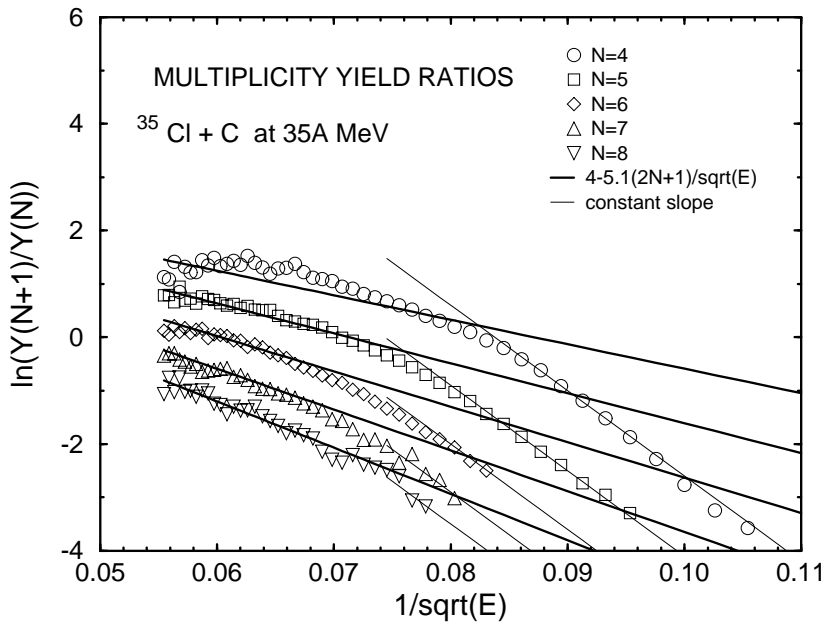}
\vspace{1.0 cm}
\caption{\protect\footnotesize Natural logarithm of multiplicity yield
ratios  for \clv\ + \cii\ at 35$A$ MeV, plotted as a function of
$E^{-1/2}$. Heavy lines represent ratios based on ``barrier-independent''
calculations of $N$ and $\sigma$; light lines represent constant slope.}
\end{figure}

\section{Comparison of Derived and Measured Statistical Observables}

Fig. 3 shows the kinetic energy fraction, $R$, plotted as a function of
excitation energy for events in which two to twelve charged products were
detected. The excitation energy attributed to each point is the mean for a
given multiplicity. Filled circles are the centroids of the $R$
distributions, with bars indicating the distribution widths, $\sigma_R$.
The $<R>=0.50$ value and associated widths from equation (7) are
indicated by the dashed curves. To investigate the effects of a more
comprehensive statistical treatment, a calculation was performed with the
code, GEMINI \mycite{char}, with average gamma-ray emission energies
obtained from a modified version of PACE \mycite{gavr}, and the results
for three excitation energies plotted as open circles. The data are
obviously in good agreement  with both the schematic calculation and the
statistical code results, though some bias due to detector acceptance may
be apparent at the lowest multiplicities. The statistical nature of the
decay is reflected in the $R$ distributions over the range of excitation
energies from 2 to nearly 7 MeV per nucleon. The narrowness of the
distribution supports the prediction of equation (7) and indicates that,
even for relatively low multiplicities, $R$ may be useful as an
event-by-event indicator of statistical decay. Indeed, the technique has
recently been applied to a series of projectile fragmentation
experiments\mycite{rroy}, giving $<R>$ values from 0.50 to 0.57.

The mean number of charged fragments emitted should, according to
equation (11), be proportional to the square root of the excitation
energy. Fig. 4 shows the centroids and widths of the measured
multiplicity distributions as a function of excitation energy. The solid
line represents a square-root proportionality, and the widths determined
by equation (12) are indicated by the dashed lines. The centroids and
widths predicted by our schematic statistical derivation are in good
agreement with experiment for the higher excitation energies, but at
lower energies, where the Coulomb barrier might impose upper limits on
the number of charged particles, the experimental widths are narrower.

Moretto {\em et al.}\mycite{mor1} have demonstrated that for IMF
emission, which may be dominated by barrier effects, the natural
logarithm of the multiplicity yield ratio is a linear function of
$E^{-1/2}$. We plot this ratio for emission of {\em all} ions in Fig.5.
Note that the data do not, except at the lowest excitation energies, have
the linear dependence upon $E^{-1/2}$ expected from the systematics of
ref. 5. Based on the behavior of the widths in Fig. 4, it would instead
be reasonable to expect agreement at higher excitations with multiplicity
yield ratios predicted for barrier-independent emission. In this case,
the assumption of gaussian multiplicity distributions gives
\begin{equation}
\ln (\frac{Y(N+1)}{Y(N)})=4-\frac{2(2N+1)}{k\sqrt{E}},
\end{equation}
where $k$ is the proportionality constant of equation (11). These
``statistical'' ratios, indicated by the heavy lines in Fig. 5, do, in
fact, approximate the data at high excitation energies. For lower
excitation energies, the constant slopes of the barrier-dominated
systematics may be more appropriate.

\section{Conclusions}

\begin{itemize}
\item A schematic calculation showed, for $E \gg B$, that the partition
of decay energy between kinetic energy of emission and $Q_0$
should be in a constant ratio for statistical decay.
\item The kinetic
energy fraction, $R = \Sigma E_{kin}/(\Sigma E_{kin} + \Sigma Q_0),$ and
its width were measured for a well-determined reaction mechanism.
\item The predicted
mean and distribution width for $R$ were observed in the data and
reproduced in calculations with a well-known statistical decay code.
\item At sub-vaporization excitation energies, emission at the barrier
has a different energy dependence than emission well above the barrier.
\item The relationship between the two processes is exemplified by many
of the correlations observed between charged-particle production, IMF
production, and excitation energy.
\end{itemize}


\begin{thebibliography}{99}

\bibitem{bowm}
D.R. Bowman {\em et al.}, Phys. Rev. {\bf C46} (1992) 1834.

\bibitem{hube}
J. Hubele {\em et al.}, Z. Phys {\bf A340} (1991) 263.

\bibitem{phai}
L. Phair, private communication.

\bibitem{pori}
N. Porile, Proc. XI Winter Workshop on Nuclear Dynamics, Key West,
Florida, 1995 Feb 11-17.

\bibitem{mor1}
L.G. Moretto, D.N. Delis, and G.J. Wozniak, Phys. Rev. Lett. {\bf 71}
(1993) 3935.

\bibitem{mor2}
L.G. Moretto {\em et al.}, Phys. Rev. Lett.
{\bf 74} (1995) 1530.

\bibitem{morr}
P. Morrison in {\em Experimental Nuclear Physics, Vol. II}, ed.
E. Segr\'{e} (John Wiley and Sons, New York, 1953), p.173.

\bibitem{siem}
P.J. Siemens {\em et al.}, Phys. Lett. {\bf 36B} (1971) 24.

\bibitem{pru1} C. Pruneau {\em et al.}, Nucl. Inst. and Meth. in Phys.
Res. {\bf A297} (1990) 404.

\bibitem{hor1}
D. Horn {\em et al.}, Nucl. Inst. and Meth. in Phys. Res.
{\bf A320} (1992) 273.

\bibitem{colo}
N. Colonna {\em et al.}, Phys. Rev. Lett. {\bf 62} (1989) 1833.

\bibitem{hor2} D. Horn {\em et al.}, PR-TASCC-4: 3.1.7; AECL-10545.

\bibitem{hor3}
D. Horn {\em et al.}, Proc. Int. Workshop on Heavy-Ion Fusion, Padova,
Italy, 1994 May 24-27, editors A.M. Stefanini {\em et al.} (World
Scientific, 1994) 208.

\bibitem{char} GEMINI code: R.J. Charity {\em et al.}, Nucl. Phys.
{\bf A483}(1988)371.

\bibitem{gavr} PACE2 code: A. Gavron, Phys. Rev. {\bf C21} (1980) 230,
modified by J.R. Beene.

\bibitem{rroy}
R. Roy {\em et al.}, to be published in Proc XXXIII International Winter
Meeting on Nuclear Physics, Bormio, Italy, 1995 Jan 23-28.

\end{thebibliography}
\end{document}